\documentstyle[prl,aps,twocolumn,epsf,floats]{revtex}

\newcommand{\beq}{\begin{equation}}
\newcommand{\eeq}{\end{equation}}
\newcommand{\bea}{\begin{eqnarray}}
\newcommand{\eea}{\end{eqnarray}}

\newcommand{\bs}{{\bf{S}}}
\newcommand{\bB}{{\bf{B}}}
\newcommand{\bL}{{\bf{L}}}
\newcommand{\bM}{{\bf{M}}}

\newcommand{\etal}{{\em et al.}}

\def\jour#1#2#3#4{{#1} {\bf #2}, #3 (#4).}
\def\tit#1#2#3#4#5{{#1} {\bf #2}, #3 (#4).}

\def\prl{Phys. Rev. Lett.}

\def\prb{Phys. Rev. B}
\def\jpco{J. Phys. Cond. Mat}

\def\zpb{Z. Phys. B}

\begin{document}
\draft

\twocolumn[\hsize\textwidth\columnwidth\hsize\csname @twocolumnfalse\endcsname

\title{Magnetic susceptibility of  diluted pyrochlore and SCGO antiferromagnets }
\author{R. Moessner$^1$\ and A. J. Berlinsky$^2$}
\address{$^1$Department of Physics, Jadwin Hall, Princeton University, Princeton, NJ 08544, USA\\
	$^2$Department of Physics and Astronomy, McMaster University, Hamilton, Ontario, Canada L8S 4M1}
\date{\today}

\maketitle

\begin{abstract}
We investigate the magnetic susceptibility of the classical Heisenberg
antiferromagnet with nearest-neighbour interactions on the
geometrically frustrated pyrochlore lattice, for a pure system and in
the presence of dilution with nonmagnetic ions. Using the fact that
the correlation length in this system for small dilution is always
short, we obtain an approximate but accurate expression for the
magnetic susceptibility at all temperatures. We extend this theory to
the compound $SrCr_{9-9x}Ga_{3+9x}O_{19}$\ ($SCGO$) and provide an
explanation of the phenomenological model recently proposed by
Schiffer and Daruka (Phys.\ Rev.\ B {\bf 56}, 13712 (1997)).
\end{abstract}


]

\begin{figure}
\epsfxsize=3.1in
\centerline{\epsffile{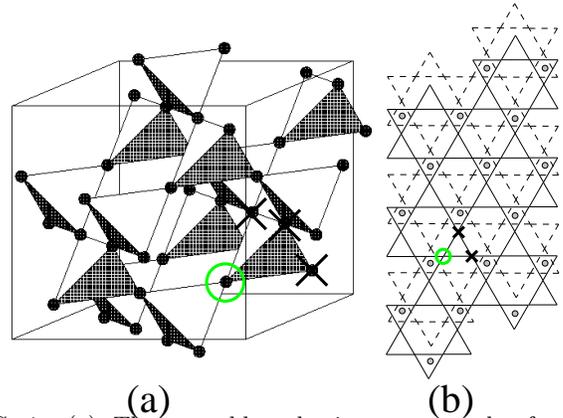}}
\caption{(a) The pyrochlore lattice, a network of corner-sharing
tetrahedra. (b) The trilayer $SCGO$ lattice, consisting of
corner-sharing tetrahedra and triangles.  The top and bottom kagome
layers are denoted by solid and broken lines, respectively, and the
intervening triangular lattice by circles.  By removing the spins
labelled by crosses, only the encircled site of the $q=1$ unit thus
generated is left occupied.}
\label{fig:scgopyro}
\end{figure}

The study of magnetic systems with competing interactions has
uncovered a wide variety of different physical phenomena\cite{diep},
such as glass transitions and the existence low-temperature disordered
phases.  In a class of problems, which has received a great deal of
attention recently\cite{reviews}, competition between
nearest-neighbour antiferromagnetic exchange interactions arises due
to the fact that, in a group of $q\geq 3$\ interacting spins, it is
not possible for each spin to be antialigned with all its $q-1$\
neighbours.  Geometric frustration of this kind can give rise to
macroscopic ground-state degeneracies, and as early as 1956, a type of
lattice was identified\cite{andersonpyro} for which the ground-state
degeneracy is particularly large. In this kind of lattice, the
frustrated units (triangles for the kagome and tetrahedra for the
pyrochlore lattice, Fig.~\ref{fig:scgopyro}) are arranged to share
sites (neighbouring units share one spin) instead of bonds, as is the
case in the more familiar triangular and face-centred cubic lattices.

Since the manifold of ground states does not provide an intrinsic
energy scale, any perturbation to the simple nearest-neighbour
exchange Hamiltonian has to be considered strong and can potentially
select different low-temperature physics from the vast range of
possibilities provided by the macroscopic degeneracy. One specatcular
example is the recently experimentally discovered magnetic analogue of
ice \cite{andersonpyro,spinice,iceth}.  The recent surge in
theoretical research on these systems results in large part from
experimental developments (see, e.g.,
Refs.~\cite{spinice,expts,tbti}). Currently, a systematic study of
compounds is under way, in which different rare earth and transition
metal ions are placed on the pyrochlore lattice, each of which comes
with its own peculiar properties (such as anisotropic or longer-range
interactions), so that the space of possible Hamiltonians is mapped
out increasingly well.

In this Letter, we study the magnetic susceptibility, $\chi$, of a
classical pyrochlore Heisenberg antiferromagnet as a function of
temperature, both for the pure system and in the presence of
disorder. The magnetic susceptibility is a particularly interesting
quantity, since for a large class of geometrically frustrated magnets,
its inverse surprisingly stays close to the linear Curie-Weiss law
down to temperatures {\em much lower} than the Curie-Weiss
temperature, $\Theta_{CW}$, where mean-field theory predicts a
transition \cite{reviews}. (Eventually, at $T_F\ll\Theta_{CW}$, such
magnets usually, but not always \cite{tbti}, freeze
\cite{gingrascrit}, a feature which is absent from the simple
classical pyrochlore Heisenberg model
\cite{villain,reimersmc,martinmc,pyroshlo}.) However, to our
knowledge, no analytical expression, exact or approximate, which is
valid at all temperatures, has so far been proposed for pyrochlore or
$SCGO$\ antiferromagnets (Fig.~\ref{fig:scgopyro}), although for the
kagome lattice, there has been some work at T=0\cite{shender}, for
infinite-component spins\cite{garanin}, and using a high-order
high-temperature series expansion\cite{harrisab}.

Although the presence of at least a small amount of disorder is
inevitable, its effect on pyrochlore magnets has so far only been
studied qualitatively\cite{villain,pyroshlo}; again there has been
some work for the kagome lattice at $T=0$\cite{kagomedis}.
Recently, Schiffer and Daruka \cite{schifferdaruka} have made the
intriguing observation, described in detail below, that the deviation
from linear Curie-Weiss behaviour at low temperatures for strongly
frustrated magnets appears almost universally as a downturn of
$\chi^{-1}$, and that this tendency becomes stronger as the dilution
is increased. This effect still awaits a theoretical explanation.

We present here a theory which provides an expression for the energy
and susceptibility of a classical Heisenberg pyrochlore
antiferromagnet which is exact at zero temperature and asymptotically
correct for large temperatures, being a good approximation in
between. Our treatment of the effect of dilution with non-magnetic
ions applies in the limit of low disorder and is shown by Monte Carlo
(MC) simulations to give reliable results for dilutions as large as
20\%. We explain the empirical findings of Ref.~\cite{schifferdaruka}
for $SCGO$\ by identifying two relevant mechanisms responsible for the
downturn of $\chi^{-1}$, one due to dilution, the other present even
in a pure system.

We start with the fundamental observation that the spin-spin
correlations of a pure classical Heisenberg antiferromagnet on the
pyrochlore lattice are always short-ranged
\cite{villain,reimersmc,martinmc,pyroshlo}. A small amount of dilution
does not affect this property and induces neither ordering nor
glassiness \cite{villain}. This ceases to be the case, at the latest,
when 1/4 of all sites are vacant.  At this dilution, the possibility
of moving local clusters of spins at no cost in energy is lost as the
ground-state degeneracy of the magnet ceases to be
macroscopic\cite{pyroshlo}. We make use of this property by treating
the weakly diluted lattice as an arrangement of units of $q\leq 4$\
spins, which -- given the small correlation length characteristic of
this regime -- are treated as if they were decoupled.

For this approximation, which we refer to as the single-unit
approximation, to be useful, we have to know the properties of the
individual units.  These can be obtained exactly using a
Hubbard-Stratonovich transformation \cite{pyroshlo}. We define the
Hamiltonian for a group of $q$\ spins $\bs_i$\ in the presence of a
magnetic field $\bB$, to be \bea H_q(\bB)=\! J \!\!\sum_{<i,j>}\!
\bs_i \cdot \bs_j\! -\!\sum_{i=1}^q \bB \cdot \bs_i \equiv \frac{J}{2}
\left ( \bL_q-\frac{\bB}{2J}\right)^2
\label{eq:hamil} 
\eea 
where we have dropped a constant in the last equality. 
The sum on $\left<i,j \right>$\ runs over all pairs. $J>0$\ is the
antiferromagnetic bond strength, and $\bL_q$\ is the total spin of the
unit. The g-factor and the Bohr magneton have been absorbed into the
definition of $\bB$. The partition function in zero field, $Z_q$, is
calculated to be
 \bea Z_q&=&\int \{d\Omega_i\}\exp(-\beta J
\bL^2/2) \nonumber \\ &=&(2 \pi \beta J)^{-3/2}(2\pi)^q
\int\!\!\!\!\int\!\!\!\!\int d^3{\bf K} \exp(-\frac{K^2}{2\beta J})(2
\frac{\sin K}{K})^q, \nonumber \eea 
where the $\{\Omega_i\}$ are the directions of the  
$\{\bs_i\}$ and $\beta=1/T$ is the inverse temperature. 
The final
integrals may be evaluated in terms of the error function Erf, and we
obtain, up to an overall constant:
 \bea Z_2(T)&=&T\Bigl(1-e^{-{2\over T}}\Bigr),\\
Z_3(T)&=&T^{3\over 2}\,
\Bigl\{3\,{\rm Erf}\Bigl(\sqrt{2/T}\Bigr)-
{\rm Erf}\Bigl(3\sqrt{2/T}\Bigr)\Bigr\},\\
Z_4(T)&=& T^{3\over 2}\,
\Bigl\{ 2\,{\rm
Erf}\Bigl(\sqrt{2/T}\Bigr) -  \,{\rm
Erf}\Bigl(\sqrt{8/T}\Bigr)\nonumber \\
&-& \sqrt{T/8\pi}\Bigl(1-e^{-{2\over T}}\Bigr)^2\Bigl(3+2e^{-{2\over T}}+
e^{-{4\over T}}\Bigr)
 \Bigr\} \eea

The susceptibility per spin, $\chi_q$, can be obtained from these
expressions via the fluctuation-dissipation theorem: $\chi_q=(\left<
M^2\right>-\left< M\right>^2)/(3Tq)$. Noting that for one unit,
$M^2=\bL^2$\ is proportional to the energy, we finally obtain
(using $\left< M\right>=0$), $\chi_q=2T/(3qJ)\frac{\delta}{\delta
T}\ln Z_q(T)$.

For the full system, the Hamiltonian is given by the sum over the
Hamiltonians of all $N$\ units, $H=\sum_{\alpha=1}^{N}
H_{q_\alpha}^{(\alpha)}(\bB/2)$; here, $q_\alpha$\ is the number of
spins in unit $\alpha$, and the magnetic field is divided by two to
avoid double counting.

\begin{figure}
\epsfxsize=3.5in
\centerline{\epsffile{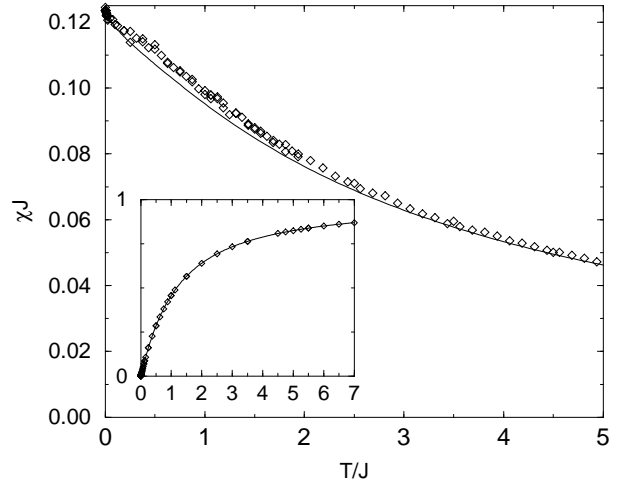}}
\caption{Susceptibility per spin, $\chi$, versus temperature from the
single-unit approximation (line) and MC simulations on 10976 spins
(diamonds).  Inset: Zero-field energy per spin versus temperature,
both in units of $J$.}
\label{fig:susc}
\end{figure}

Let us first consider the clean system, $x=0$, where all units have
$q=4$. In the inset to Fig.~\ref{fig:susc}, we plot the energy per spin versus
temperature in zero field, as obtained from the single-unit
approximation and from Monte Carlo simulations.
This plot demonstrates the quality of the approximation. The
theoretical curves in Fig.~\ref{fig:susc} are scaled overall to take
into account the fact that in the full system, each spin belongs to
two tetrahedra and has six rather than three nearest neighbours.
Fig.~\ref{fig:susc}\ also shows a plot of the susceptibility.  At high
temperature, the analytic approximation and the simulation data both
follow the Curie-Weiss law and therefore agree. As the temperature is
lowered, the analytical result is only approximately valid, since it
neglects the emerging longer-range correlations. Still, since these
are never strong, the disagreement never exceeds 5\%. Remarkably, at
the lowest temperatures, the two meet again, and at $T=0$, our theory
gives the exact result $\chi(0)=1/(8J)$.

Next, we turn to the diluted system. At high $T$, the effect of
dilution is to decrease the average number of bonds of each spin, thus
giving a reduced Curie-Weiss temperature
$\Theta_{CW}(x)=(1-x)\Theta_{CW}(0)$\cite{spalek}. The main effect of
disorder is seen by considering the susceptibility at $T=0$. On 
diluting the pyrochlore lattice, the probability of a
tetrahedron containing $q$\ spins is given by $P_q(x)=\left( {4
\atop{q}}\right) (1-x)^q x^{4-q}$. The effect of a magnetic field on
the total spin of a unit is the same for all $q\geq2$.  From
Eq.~\ref{eq:hamil}, we can read off that $H_q({\bf B})$\ is minimised by choosing
$\bL_q=\bB/(2J)$, which is possible for all magnetic fields of
strength $B\leq 2Jq$. A difference in the magnetic susceptibility per
spin arises only because the total magnetisation is shared between
a total of $q<4$\ spins in a diluted unit.  In any case, the
susceptibility is finite since $\bL_q\propto\bB$. This is in striking
contrast to the case of $q=1$, where at $T=0$  an infitesimal
field suffices to align the spin, and hence $\chi_1(0)$\ is infinite.

In the full system at low dilution, a unit with $q=1$\ in general
corresponds not to an isolated spin but to one which also belongs to a
unit with $q\geq2$ (see Fig.~\ref{fig:scgopyro}). In this regime,
there remain sufficiently many undiluted tetrahedra to generate
a macroscopically degenerate ground state. This in turn provides the
possibility of reorienting local spin clusters at no cost in energy
\cite{pyroshlo}.
As a result, the correlation length remains short at all temperatures,
and the spins in the $q=1$\ units, which are
well-separated at low dilution, behave like {\em paramagnetic} spins
even though they are not isolated.

\begin{figure}
\epsfxsize=2in
\centerline{\epsffile{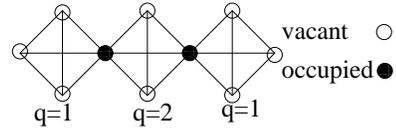}}
\caption{A configuration containing two $q=1$ units but with vanishing
magnetisation at low $T$.}
\label{fig:reduce}
\end{figure}

The behaviour at low temperatures is most easily discussed in terms of
the thermally averaged square of the total magnetisation of the
system, $\left< M^2(x,T)\right>$. By equipartition, a unit with $q\geq
3$\ has $\bL_q^2\approx3 T/J$, and one with $q=2$\ has
$\bL_2^2\approx2 T/J$, which vanishes as $T\rightarrow 0$. By
contrast, a unit with $q=1$\ has $\bL_1^2=\bs^2\equiv1$. Since the
magnetisation of the system equals the sum of the magnetisations of
the units, we obtain $\bM=\sum_{\alpha=1}^N\bL^{(\alpha)}/2$, where
the factor 1/2 accounts for the fact that each spin belongs to two
units. Therefore,
$\left<M^2(x,T)\right>=\sum_{\alpha=1}^N(\bL^{(\alpha)})^2/4+
\sum_{\alpha\neq\alpha^\prime}
\left<\bL^{(\alpha)}\cdot\bL^{(\alpha)^\prime}\right>/4$.  We find
that $\sum_{\alpha\neq\alpha^\prime}
\left<\bL^{(\alpha)}\cdot\bL^{(\alpha)^\prime}\right>=0$\ at low
temperature, as would be appropriate for completely uncorrelated
$\bL^{(\alpha)}$. However, our simulations suggest that this sum
vanishes not because the individual terms are zero, but rather because
the terms in the sum exactly cancel for $T\rightarrow 0$. Hence,
$\left<M^2(x,T)\right>=6TN(1-x)\chi(x,T)=\sum_qN_q(x)\bL_q^2(T)/4$,
where $N_q=N P_q(x)$\ is the number of units with $q$\ spins, and
$\bL_q^2$\ is given by the single-unit expression derived above.  The
estimate $\left<M^2(0)\right>=N_{1}/4=Nx^3(1-x)$\ is asymptotically
exact in the limit $x\rightarrow 0$. For increasing dilution, terms
higher order in $x$\ change the situation. For example, the
configuration depicted in Fig.~\ref{fig:reduce}, which is improbable
at small $x$, contains two $q=1$\ units but has a vanishing
magnetisation at $T=0$.

\begin{figure}
\epsfxsize=3.5in
\centerline{\epsffile{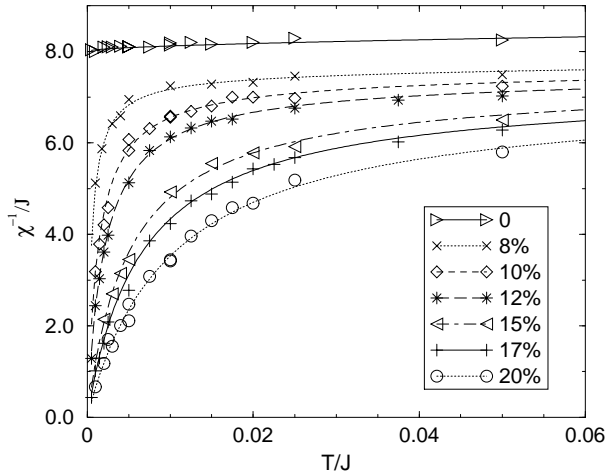}}
\caption{The inverse susceptibility for different dilutions $x$. The
lines are given by the single-unit approximation without any
adjustable parameters. Data points are from MC simulations of a
lattice of 5488 tetrahedra.}
\label{fig:dilsusc}
\end{figure}

In Fig.~\ref{fig:dilsusc}, we plot the inverse susceptibility
$\chi(T)^{-1}$\ for different dilutions in the range $0\leq x\leq0.2$\
at low temperature.  Note that for
$x$\ as large as 0.2, the agreement of our theory with the MC
simulations is excellent.


It is worth emphasizing that at any nonzero dilution, the low
temperature susceptibility is dominated by the $q=1$\ units.  This is
analogous to a Griffiths-McCoy effect as it arises from the rare,
local event of three neighbouring sites being vacant.  The temperature
$\tilde{T}$\ below which the paramagnetic regions dominate can be
defined as the point where the magnetisation due to the $q=1$\ units
equals that of the other units combined, which gives $\tilde{T}\propto
x^3$. This is why the temperature at which the downturn of
$\chi^{-1}$\ becomes visible is small and increases with disorder.

Finally, we address in detail the work by Schiffer and
Daruka \cite{schifferdaruka}, who
proposed and successfully used a two-population model to fit the
measured $\chi$ to a form $\chi=
\frac{C_1}{T+\Theta_{w1}}+\frac{C_2}{T+\Theta_{w2}}$. Here $C_1$\ and
$\Theta_{w1}$\ are the Curie constant and Curie-Weiss temperature of a
`correlated' population which forms momentless clusters as
$T\rightarrow 0$, while $C_2$\ and $\Theta_{w2}$\ are  the
 parameters for an `orphan' population, which was surmised to
be excluded from the correlated population.
Analysing experimental data on $SCGO$\ from Refs.\cite{scgosusc}, they
found that $\Theta_{w2}$\ can be set to zero, so that the orphan
population in this case appears truly paramagnetic. For this compound,
they present a series of different dilutions, independently determined
to be in the range $0.11\leq x\leq 0.61$, and find that
$\Theta_{w1}(x)=(1-x)\Theta_{w1}(0)$ and $C_2(x)\propto x$.

The model we have presented above for the pyrochlores also applies to
$SCGO$ \cite{pyroshlo}, which consists of two kagome layers connected
by an intervening triangular layer, and can be thought of as a slab of
pyrochlore cut in a $\left<111\right>$-direction
(Fig.~\ref{fig:scgopyro}). Every other kagome triangle is associated
with a site in the triangular layer, and these together form $q=4$\
units which generate the short correlation length required for our
theory \cite{pyroshlo}.

The two main differences for a theory of $SCGO$\ are as
follows. Firstly, only two spins (in a triangle not associated with a
site of the triangular lattice, as shown in Fig.~\ref{fig:scgopyro}b)
have to be removed to generate a $q=1$\ unit. Hence these units occur
with a probability $3x^2/2$\ at low dilution rather than $4x^3$\ as in
the pyrochlore. Secondly, an additional mechanism is present even for
an undiluted system, which generates a paramagnetic response in
$SCGO$. If the interlayer coupling, which acts between spins of the
triangular and the kagome layers, has strength $J^{\prime}$\ different
from $J$, the kagome intralayer coupling, the Hamiltonian for a $q=4$\
unit reads $H_4(\bB)=(J/2) \left\{ \left(
\bL^\prime-\bB/(2J)\right)^2-(1-{J^\prime}/{J})\bB\cdot\bs_t\right\}$,
where $\bs_t$\ is the spin in the triangular layer, and
$\bL^\prime=(J^\prime/J)\bs_t+\sum_{q=1}^{3}\bs_i$. In the
zero-field ground state, each $q=4$\ unit has a finite magnetisation
of magnitude $1-J^\prime/J$. These individual magnetisation vectors
are aligned at $T=0$ even by an infinitesimal field, yielding an
infinite susceptibility. Arguments within the single-unit
approximation along the lines of those given above predict a
susceptibility that diverges as $1/T$\ at low temperatures. The value
of $J^\prime/J$\ is not known accurately, and there is no reason for
it to be exactly 1.  For a system known to be clean, the divergence of
the susceptibility could be used to estimate this quantity.

As a result, at zero and at low dilutions, the model of
Ref.~\onlinecite{schifferdaruka}\ is reasonable even though the orphan
population is not excluded from the correlated one but present even
for the pure system and enhanced by the creation of $q=1$ units at
small dilution. (Genuinely isolated spins of course appear at high
dilution.) The empirical fit $C_2(x) \propto x$\ is clearly at
variance with our theory, since it yields
$C_2(x)=(1-J^\prime/J)^2/21+x^2/14$ for small $x$.  However, there is no
contradiction, since the experimental data consists of only five
points, and there are no data points sufficiently close to $x=0$ to
distinguish between the two functional forms of $C_2$.

A quantitative comparison of our theory with experiment may be
possible in the pyrochlores, since some compounds can be grown with
very small amounts of disorder \cite{gaurei}. Introducing
controlled amounts of dilution with nonmagnetic ions can then produce
a series of samples allowing measurement of susceptibility as a
function of disorder and temperature.

One of us (R.M.) acknowledges useful discussions with J. Chalker,
M. Gingras, C. Henley and P. Schiffer. This work was supported in
part by grants from the Deutsche Forschungsgemeinschaft and the
Natural Sciences and Engineering Research Council of Canada.


\begin{references}

\bibitem{diep}
For a collection of articles, see {\em Magnetic systems with competing interactions : frustrated spin systems}, edited by H. T. Diep (World Scientific, Singapore, 1994).                           

\bibitem{reviews}
For reviews, see:  
A. P. Ramirez, Annu. Rev. Mater. Sci. {\bf 24},
453, (1994);  
P. Schiffer and A. P. Ramirez, Comments
Cond.\ Mat.\ Phys.\ {\bf 18}, 21 (1996); and
M. J. Harris and M. P. Zinkin,
 \jour{Mod.\ Phys.\ Lett.\ B}{10}{417}{1996}

\bibitem{andersonpyro}
P. W. Anderson, \tit{Phys. Rev.}{102}{1008}{1956}{Ordering and
Antiferromagnetism in Ferrites}

\bibitem{spinice}
M. J. Harris \etal,
      \jour{\prl}{79}{2554}{1997} 
A. P. Ramirez \etal, to appear in Nature. 

\bibitem{iceth} R. Moessner, \tit{\prb}{57}{R5587}{1998}{Relief and
generation of frustration in pyrochlore magnets by single-ion
anisotropy} 
S. T. Bramwell and M. J. Harris,
\tit{\jpco}{10}{L215}{1998}{Frustration in Ising-type spin models on
the pyrochlore lattice} 
R. Siddharthan \etal, cond-mat/9902010

\bibitem{expts}
for example, S. R. Dunsiger \etal,
\tit{\prb}{54}{9019}{1996}{Muon spin relaxation investigation of the
      spin dynamics of geometrically frustrated antiferromagnets
      $Y_2Mo_2O_7$\ and $Tb_2Mo_2O_7$} 
R. Ballou, E. Leli\`evre-Berna
      and B. Fak, \tit{\prl}{76}{2125}{1996}{Spin Fluctuations in
      (Y_{0.97}Sc_{0.03}Mn_2: A Geometrically Frustrated, Nearly
      Antiferromagnetic, Itinerant Electron System} 
Y. Shimakawa \etal,
      \tit{\prb}{59}{1249}{1999}{Crystal structure, magnetic and
      transport properties, and electronic band structure of A2Mn2O7
      pyrochlores (A = Y, In, Lu, and Tl)} 

\bibitem{tbti}
J. S. Gardner \etal, \tit{\prl}{82}{1012}{1999}{Cooperative
paramagnetism in the geometrically frustrated pyrochlore
antiferromagnet Tb2Ti2O7}

\bibitem{gingrascrit}
M. J. P. Gingras \etal, {\prl} {\bf 78}, {947} (1997) and references
therein. 

\bibitem{villain}
J. Villain, \tit{\zpb}{33}{31}{1979}{Insulating Spin Glasses}

\bibitem{reimersmc}
J. N. Reimers, \tit{\prb}{45}{7287}{1992}{Absence of long-range order
in a three-dimensional geometrically frustrated antiferromagnet}

\bibitem{martinmc} %
M. P. Zinkin, M. J. Harris and T. Zeiske,
\tit{\prb}{56}{11786}{1997}{Short-range magnetic order in the
frustrated pyrochlore antiferromagnet $CsNiCrF_6$}

\bibitem{pyroshlo} 
R. Moessner and J. T. Chalker,
\tit{\prl}{80}{2929}{1998}{Properties of a classical spin liquid: The
Heisenberg pyrochlore antiferromagnet}
\tit{\prb}{58}{12049}{1998}{Low-temperature properties of classical
geometrically frustrated antiferromagnets}


\bibitem{shender} 
E. F. Shender, and P. C. W. Holdsworth, {\em Order
by disorder and topology in frustrated magnetic systems} in
M. M. Millonas (ed.) {\em Fluctuations and Order: The New Synthesis}
(MIT Press, Boston, 1994).

\bibitem{garanin}
D. A. Garanin and B. Canals, \tit{\prb}{59}{443}{1999}{Classical spin liquid: Exact solution for the infinite-component antiferromagnetic model on the kagome lattice}

\bibitem{harrisab}
A. B. Harris, C. Kallin and A. J. Berlinsky,
\tit{\prb}{45}{2899}{1992}{Possible Neel orderings of the Kagome
antiferromagnet}

\bibitem{kagomedis}
E. F. Shender \etal, 
\tit{\prl}{70}{3812}{1993}{Kagome Antiferromagnet
with Defects: Satisfaction, Frustration, and Spin Folding in a Random
Spin System} D. L. Huber and W. Y. Ching, \tit{\prb}{47}{3220}{1993}{ANTIFERROMAGNETIC MAGNONS IN DILUTED TRIANGULAR AND KAGOME LATTICES}


\bibitem{schifferdaruka}
P. Schiffer and I. Daruka, \tit{\prb}{56}{13712}{1997}{Two-population model for anomalous low-temperature magnetism in geometrically frustrated magnets}

\bibitem{spalek}
J. Spalek \etal, \tit{\prb}{33}{3407}{1986}{MAGNETIC-SUSCEPTIBILITY OF SEMIMAGNETIC SEMICONDUCTORS - THE HIGH-
      TEMPERATURE REGIME AND THE ROLE OF SUPEREXCHANGE}

\bibitem{scgosusc} 
A. P. Ramirez \etal, \jour{\prl}{64}{2070}{1990}
\jour{\prb}{45}{2505}{1992} P. Schiffer \etal,
\jour{\prl}{77}{2085}{1996}

\bibitem{gaurei}%
B. D. Gaulin \etal, 
      \tit{\prl}{69}{3244}{1992}{Spin freezing in the geometrically
      frustrated pyrochlore antiferromagnet $Tb_2Mo_2O_7$}

\end{references}
\end{document}